\documentclass[aps,prb,twocolumn,superscriptaddress,nofootinbib,floatfix]{revtex4-2}

\usepackage{graphicx}%
\usepackage{amsmath,amssymb,amsfonts}%
\usepackage{mathrsfs}%
\usepackage{bm}%
\usepackage{dcolumn}%
\usepackage{braket}%
\usepackage[normalem]{ulem}%
\usepackage{verbatim}
\usepackage{textcomp}%
\usepackage{xcolor}%
\usepackage{booktabs}%

\newcommand{\lambdaWRes}[0]{\lambda_{\mathrm{W}}}
\newcommand{\etaLD}[0]{\lambda_{\mathrm{L}}}

\newcommand{\Diff}[2]{\frac{ \partial{#1} }{ \partial{#2} }}
\newcommand{\DDiff}[2]{\frac{ \partial^2{#1} }{ \partial{#2}^2 }}

\newcommand{\Difft}[2]{\frac{ \mathrm{d}{#1} }{ \mathrm{d}{#2} }}

\newcommand{\kv}[0]{\mathbf{k}}

\newcommand{\rv}[0]{\mathbf{r}}
\newcommand{\pv}[0]{\mathbf{p}}

\newcommand{\Sv}[0]{\mathbf{S}}

\newcommand{\psiv}[0]{\mathbf{\Psi}}

\newcommand{\rmd}[1]{\mathrm{d}{#1}}

\newcommand{\me}[0]{m_\mathrm{e}}

\newcommand{\epso}[0]{\varepsilon_0}
\newcommand{\epstat}[0]{\varepsilon_\mathrm{s}}
\newcommand{\epsinf}[0]{\varepsilon_\infty}

\newcommand{\gammaD}[0]{\gamma_\mathrm{D}}

\newcommand{\nion}[0]{n_{+}}

\newcommand{\EC}[0]{E_\mathrm{c}}
\newcommand{\Ep}[0]{E^{\prime}}

\newcommand{\Evb}[0]{\mathbf{E}_\mathrm{b}}

\newcommand{\Eg}[0]{E_\mathrm{g}}
\newcommand{\Deltasp}[0]{\Delta_\mathrm{so}}
\newcommand{\mestar}[0]{m_0^{*}}

\newcommand{\omegaP}[0]{\omega_\mathrm{p}}

\newcommand{\fFD}[0]{f_\mathrm{FD}}

\newcommand{\zetaP}[0]{\frac{\partial\zeta}{\partial n}}

\newcommand{\ii}[0]{{i^{\prime}}}
\newcommand{\jj}[0]{{j^{\prime}}}

\newcommand{\ThetaMat}[0]{\widehat{\Theta}}

\newcommand{\GammaMat}[0]{\widehat{\Gamma}}

\newcommand{\mResp}[0]{m^{*}}
\newcommand{\un}[0]{\rho }

\newcommand{\trans}[0]{\mathrm{T}}

\begin{document}
\raggedbottom
\newcommand{\backhead}[1]{\par\vspace{14pt}\noindent{\bfseries\MakeUppercase{#1}}\par\nopagebreak\vspace{4pt}}

\title{Generalized Lorentz Framework for Active Semiconductor Quantum Surface Plasmonics}

\author{Wancong Li}
\thanks{W. Li and Q. Zhou contributed equally to this work.}
\affiliation{School of Physics and Electronics, Hunan University, Changsha 410082, China}
\affiliation{School of Physics, Institute for Quantum Science and Engineering, Hubei Fundamental Research Center for Physics and Hubei Key Laboratory of Gravitation and Quantum Physics, Huazhong University of Science and Technology, Wuhan 430074, China}

\author{Qiang Zhou}
\affiliation{School of Physics, Institute for Quantum Science and Engineering, Hubei Fundamental Research Center for Physics and Hubei Key Laboratory of Gravitation and Quantum Physics, Huazhong University of Science and Technology, Wuhan 430074, China}

\author{Runzhi Zou}
\affiliation{School of Physics,Institute for Quantum Science and Engineering, Hubei Fundamental Research Center for Physics and Hubei Key Laboratory of Gravitation and Quantum Physics, Huazhong University of Science and Technology, Wuhan 430074, China}

\author{Zi He}
\affiliation{School of Microelectronics, Nanjing University of Science and Technology, Nanjing 210094, China}

\author{Pu Zhang}
\email{puzhang0702@hust.edu.cn}
\affiliation{School of Physics, Institute for Quantum Science and Engineering, Hubei Fundamental Research Center for Physics and Hubei Key Laboratory of Gravitation and Quantum Physics, Huazhong University of Science and Technology, Wuhan 430074, China}

\author{Hongwei Han}
\affiliation{Michael Gr\"atzel Center for Mesoscopic Solar Cells, Wuhan National Laboratory for Optoelectronics, Huazhong University of Science and Technology, Wuhan 430074, China}

\author{Shuangchun Wen}
\affiliation{School of Physics and Electronics, Hunan University, Changsha 410082, China}

\author{Xue-Wen Chen}
\email{xuewen\_chen@hust.edu.cn}
\affiliation{School of Physics, Institute for Quantum Science and Engineering, Hubei Fundamental Research Center for Physics and Hubei Key Laboratory of Gravitation and Quantum Physics, Huazhong University of Science and Technology, Wuhan 430074, China}

\begin{abstract}
Nanostructured semiconductors provide an actively tunable platform for light-matter interactions across a broad spectral range. As feature sizes shrink to the deep nanoscale ($\sim10$ nm scale), nonclassical carrier dynamics and surface plasmonic responses become prominent, creating substantial challenges for a rigorous and efficient theoretical treatment. Here we develop a generalized Lorentz framework for quantum surface plasmonics in nanostructured semiconductors with arbitrary energy-momentum dispersion relations. The framework enables efficient quantitative treatment of carrier nonlocality, carrier surface accumulation and depletion, and nonlocal damping in three-dimensional semiconductor nanostructures, under electrical tuning and across a wide range of doping densities. Building on this framework, we further establish a rigorous quasinormal-mode analysis that provides transparent physical insight into complex surface plasmonic responses of multiscale semiconductor structures. We demonstrate electrically controlled deep-nanoscale carrier accumulation and depletion in a semiconductor nanoantenna, leading to pronounced plasmonic-response modulation that can be interpreted in terms of eigenmode shifts, creation, and annihilation. By quantifying and linking complex nonclassical carrier dynamics with plasmonic eigenmodes, our work establishes a comprehensive quantitative framework for active semiconductor quantum surface plasmonics and will foster the convergence of electronics and photonics at the deep nanoscale.
\end{abstract}

\maketitle

\section{Introduction}
Semiconductors underpin modern nanoelectronics and also offer a versatile platform for light-matter interactions across the mid- to near-infrared spectrum through plasmonic resonances arising from collective oscillations of charge carriers \cite{luther_localized_2011,naik_alternative_2013,taliercio_semiconductor_2019}. Unlike metals, semiconductors exhibit plasmonic properties actively tunable through doping and electrical gating \cite{feigenbaum_unity-order_2010,faucheaux_plasmon_2014,DeLuca2022}, offering appealing opportunities for on-chip electro-optical modulation, ultrasensitive biosensors, nonlinear effects, and all-optical switching \cite{liu_high-speed_2004,krasavin_photonic_2012,ma_indium-tin-oxide_2015,law_all-semiconductor_2013,pradhan_extreme_2014,alam_large_2016,hu_modulating_2025}. Recent advances in nanofabrication technology have enabled three-dimensional (3D) nanostructuring of semiconductors with single-digit-nanometer precision \cite{manfrinato_aberration-corrected_2017,oh_guided_2023,zhang_advances_2022}. Such capability allows precise control of the interactions among light field, charge carriers, and other fields at the deep nanoscale, fostering nanoelectronic-nanophotonic convergence. This creates a platform that hosts rich physics at the intersection of electromagnetism, quantum mechanics, and condensed-matter physics, which necessitate a sophisticated theoretical description. 
While preliminary studies have demonstrated the importance of quantum effects on carriers in one-dimensional (1D) systems \cite{abe_evolution_2002,liu_electrical_2017,Vasanelli2020}, a rigorous quantitative theoretical treatment of the carrier dynamics and surface plasmonic responses in multiscale, 3D nanostructures remain a substantial open challenge.

\begin{figure}[!t]
\centering
\includegraphics[width=\linewidth]{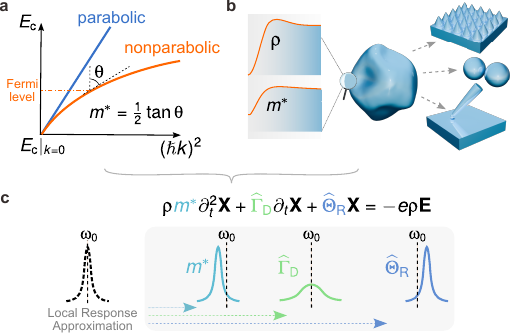}
\caption{\textbf{Key ingredients of the generalized Lorentz framework.} \textbf{a}, Illustration of nonparabolic energy-momentum dispersion for conduction electrons in $n$-doped semiconductors. \textbf{b}, Schematics of position-dependent electron density ($\rho$) and effective mass ($m^*$) near the interface of an arbitrarily shaped nanostructure. \textbf{c}, Graphic illustration of the generalized Lorentz model for driven conduction electrons.}
\label{fig:fig1}
\end{figure}

As feature sizes of doped semiconductors shrink to the deep nanoscale, nonclassical effects, such as carrier nonlocality, band bending near interfaces, and nonlocal damping with hot carrier generation (sometimes also refered as Landau damping) \cite{Aupiais2023,Yin2013,zandi_impacts_2018,khurgin_hot-electron_2024,Muravitskaya2024,gurung_control_2025}, become prominent. Electron nonlocality and nonlocal damping had been recognized as major challenges for describing quantum plasmonic phenomena in metallic nanostructures \cite{scholl_quantum_2012,ciraci_probing_2012,esteban_bridging_2012,boroviks_extremely_2022NatCommun,monticone_nonlocality_2025}. Over the past decade, these effects have been successfully treated using the d-parameter approach \cite{Christensen2017,yang_general_2019,Mortensen:2021,zhou_quasinormal_2022,PADG2023,babaze_dispersive_2023} or the self-consistent hydrodynamic models (HDM) \cite{Hongxing2015ResonanceShifts,zhang_quantum-hydrodynamic_2025,cirac2017CurrentDependent,Zhou2021,zhou_quantum_2023}. Although the HDM appears viable for semiconductors, an appropriate formulation of its central ingredient, i.e., the free-energy functional, has yet to be developed. A direct adoption of the metal-form free-energy functional for semiconductors fails even to obtain correct stationary carrier distribution, prompting researchers to rely on effective Poisson theory for this purpose \cite{zandi_impacts_2018,DeLuca2022}. Unlike in metallic systems, where the overall electron densities are fixed and electrons exhibit simple parabolic dispersion, the challenges in semiconductors are substantially escalated due to complicated band structures, largely variable carrier density, and band-bending effects near surfaces. For instance, conduction electrons in $n$-doped semiconductors generally exhibit non-parabolic energy-momentum dispersion \cite{kane_band_1957,Pisarkiewicz1990}, doping-density dependent nonlocal damping, and position-dependent effective electron mass as illustrated in Fig.\,\ref{fig:fig1}a, b. The absence of a proper formulation also precludes a rigorous quasinormal mode (QNM) analysis \cite{Yan2018,kristensen_modeling_2020,both_resonant_2022,sauvan_normalization_2022,Binkowski2024.PRB}, which would provide a transparent understanding of the intricate physics.  

Here we present a generalized Lorentz framework to describe the nonclassical effects in multiscale, 3D nanostructured semiconductors with arbitrary energy-momentum dispersion relations. Specifically, we derive a generalized free energy functional for doped semiconductors with arbitrary dispersion relations, which yields self-consistent stationary electron distribution and captures nonlocal optical responses. Moreover, we introduce a viscous tensor capable of accounting for nonlocal damping in semiconductors with largely varying doping levels and inhomogeneous electron density. Building on the framework, we provide accurate evaluation of the complex carrier dynamics and the linked quantum surface plasmonic responses of semiconductor nanostructures, which are subjected to electrical field controls and orders-of-magnitude variation in doping density. Furthermore, we establish a rigorous QNM analysis with a consistent orthonormalization scheme. We theoretically demonstrate electrically induced deep nanoscale carrier accumulation and depletion in a semiconductor nanoantenna, leading to pronounced electro-plasmon modulations that can be understood with the picture of QNM shift, creation, and annihilation.

\section{Theoretical framework}
\subsection{Generalized Lorentz model}
We here consider $n$-doped semiconductors and first derive a generalized Lorentz model for the conduction electrons.
Based on the hydrodynamic description, the Hamiltonian for the conduction electrons in the semiconductor reads 
\begin{align}
\label{Eq:HamiltonianSemiclassical}
& H[n,w] = 
F[n(\mathbf{r},t; E_\mathrm{c}(k))]  
-\!\mu\!\int \!\! \mathrm{d}\mathbf{r}\,
n(\mathbf{r},t; E_\mathrm{c}(k)) +
\notag \\
& 
\!\!\!+\!\!\! \int \!\! \mathrm{d}\mathbf{r}\bigg\{ \!\frac{\left[\nabla w(\mathbf{r},t) \! +\! e\mathbf{A}(\mathbf{r},t)\right]^2}{2m^*(\mathbf{r})}
\!- \! e\phi(\mathbf{r},t) \!
\bigg\}n(\mathbf{r},t; E_\mathrm{c}(k)) ,
\end{align}
where $n(\mathbf{r},t;E_\mathrm{c}(k))$ is the number density of the electrons and depends on the nonparabolic conduction-band dispersion $E_\mathrm{c}(k)$. 
$w(\mathbf{r},t)$ is the canonical conjugate of $n(\mathbf{r},t; E_\mathrm{c}(k))$ and $\nabla w+e\mathbf{A}$ is the mechanical momentum. 
$m^*(\mathbf{r})$ is the effective mass evaluated at the Fermi energy. Near the interface, $m^*(\mathbf{r})$ becomes position dependent (Fig.\,\ref{fig:fig1}b) owing to the variation of the electron density and nonparabolicity. The first two terms in the Hamiltonian originate from the ensemble averaged internal energy \cite{Moldabekov_hydrodynamic_plasma_2018}, where $F[n(\mathbf{r},t; E_\mathrm{c}(k))]$ represents the free energy functional and incorporates the complex nonlocal interactions of electrons in nanostructured semiconductors. $\mu$ is a constant denoting the thermo-equilibrium chemical potential of the system, and is related to the doping density $n_+$ via 
\begin{equation}\label{Eq:defChemicalPotential}
n_+ = \int_0^{\infty}\!\mathrm{d}k \frac{k^2}{(\exp[\beta E_\mathrm{c}(k)-\beta\mu]+1)\pi^2}, 
\end{equation}
with $\beta$ being the thermodynamic beta (see Supplementary Note~2). 
The third term in the Hamiltonian describes the electromagnetic interaction through vector potential $\mathbf{A}(\mathbf{r},t)$ and scalar potential $\phi(\mathbf{r},t)$. They encode the optical excitation field, optical response field and electrostatic controls through charging or bias. Based on electron number conservation $\!-e\partial_tn  +  \nabla \cdot \mathbf{J}\!=\!0$, one could obtain the current as  $\mathbf{J}\allowbreak\!=\!\allowbreak-ne(\nabla w + e\mathbf{A})/m^*\!$. The optical response can be conveniently described by the electrons' displacement from equilibrium $\mathbf{X}(\mathbf{r},t)$, that is related to the current via $\mathbf{J}\!=\!-n e\partial_t\mathbf{X}$, giving 
$\mathbf{X}(\mathbf{r},t)\!=\!
\int_0^{t}\,[
\nabla w(\mathbf{r},t')
\!+\!
e\mathbf{A}(\mathbf{r},t')
]/m^*(\mathbf{r})\mathrm{d}t'$.
Thus, based on one of the Hamilton's equations
$\partial_tw\!=\!-\delta H[n,w]/\delta n$,
we obtain the following equation (see Supplementary Note~1 for the derivation):
\begin{align}
\label{Eq:ForceBalance}
m^*n\partial_t^2\mathbf{X}
= -en\mathbf{E} 
-n\nabla\frac{\delta F[n]}{\delta n}
+ \mathbf{F}_\mathrm{NL}
+ \mathbf{F}_\mathrm{D},
\end{align}
recovering Newton's second law, where 
$\mathbf{E}\allowbreak\!=\!\allowbreak
-\partial_t\mathbf{A}
\allowbreak\!-\!\allowbreak\nabla\phi$ 
is the electric field. 
$\mathbf{F}_\mathrm{NL}$ collects the nonlinear terms of $\mathbf{X}$ (Supplementary Note~1). 
$\mathbf{F}_\mathrm{D}$ is an added damping force supplementing the conservative Hamiltonian 
in Eq. \eqref{Eq:HamiltonianSemiclassical}.

We consider linear response and employ a perturbative treatment with 
\begin{align}
    n&=n_0+n_1, \\
    \mathbf{E}&=\mathbf{E}_0+\mathbf{E}_1.
\end{align}
Subscript 0/1 denotes the stationary/dynamic component. Equation \eqref{Eq:ForceBalance} can be separated into the following set of equations 
\begin{gather}
\label{Eq:GeneralizedLorentzByDisplacement}
\rho{m^*}\partial_t^2\mathbf{X}
+ \widehat{\Gamma}_\mathrm{D}\partial_t\mathbf{X}
+ \widehat{\Theta}_\mathrm{R}\mathbf{X}
= {-e\rho}\,\mathbf{E}_1,
\\
\label{Eq:GeneralizedLorentzStationary}
\left(\delta F[n]/\delta n\right)_{0}
- e\phi_0
= \mu,
\\
\label{Eq:RestoringForce}
\widehat{\Theta}_\mathrm{R}\mathbf{X}
\allowbreak\!=\!\allowbreak 
\rho\nabla\!\left(\delta F[n]/\delta n\right)_1,
\end{gather}
where $\rho=n_0/n_+$ and $(\cdots)_0$ denotes taking $n=n_0$. To obtain Eq.~\eqref{Eq:GeneralizedLorentzStationary}, we have used $-\nabla\phi_0\allowbreak\!=\!\allowbreak\mathbf{E}_0$ with $\phi_0$ being the electrostatic potential. It allows the incorporation of external electrical controls such as charging and electrostatic fields. In Eq.~\eqref{Eq:RestoringForce}, $(\cdots)_1$ denotes collecting only the linear terms of $n_1$, which is related to  $\mathbf{X}$ via $n_1
\allowbreak\!=\!\allowbreak
-\nabla\cdot(n_0\mathbf{X})$. 
As Eq.~\eqref{Eq:GeneralizedLorentzByDisplacement}, we have formally derived a generalized Lorentz model for driven conduction electrons, having $\mathbf{F}_\mathrm{D}\allowbreak\!=\!\allowbreak
-n_+{\widehat{\Gamma}_\mathrm{D}\partial_t\mathbf{X}}$ as the damping force and 
$\mathbf{F}_\mathrm{R}\allowbreak\!=\!\allowbreak
-n_+{\widehat{\Theta}_\mathrm{R}\mathbf{X}}$ as the restoring force. Here $\widehat{\Gamma}_\mathrm{D}$ 
and $\widehat{\Theta}_\mathrm{R}$ are tensorial differential operators depending on $n_0$. The remaining task is to explicitly work out the functional $F[n]$ and the damping term $\widehat{\Gamma}_\mathrm{D}\partial_t{\mathbf{X}}$.

\subsection{Free energy functional and its functional derivative}
The free energy functional lies at the heart of our framework, as it synergistically incorporates the complex nonlocal effects illustrated in Fig.~\ref{fig:fig1}. In the following, we give the key steps to obtain the free energy functional and provide the detailed derivations in Supplementary Note 2. The free energy functional is defined as $F[n]\allowbreak\!=\!\allowbreak
\int\!\mathcal{F}[n]\,\mathrm{d}\mathbf{r}$, where the free energy density $\mathcal{F}[n]$ takes the form
$\mathcal{F}[n]
\allowbreak\!=\!\allowbreak \mu n 
\allowbreak-\!\allowbreak \int_{-\infty}^{\mu}\!n(\mu')\mathrm{d}\mu'$ \cite{EntropyBook2021}.
Therein, the density $n(\mu)$ at a given chemical potential $\mu$ implicitly includes the nonparabolic dispersion relation $E_\mathrm{c}(k)$, and can be derived using the Matsubara Green's function \cite{Altland_Simons_condensed_matter_2010,tomishima_thomas-fermi_1985,coleman_introduction_2015}.
With the Dyson's equation, we carry out the derivation in terms of the Moyal expansion \cite{Moyal1949,baraff_greens_1961}. We find that the free energy consists of
the generalized Thomas-Fermi (TF), von Weizs\"{a}cker (W), and exchange-correlation (XC) energies, with
$\mathcal{F}
\allowbreak=\allowbreak\mathcal{F}_\mathrm{TF}
\allowbreak+\allowbreak\mathcal{F}_\mathrm{W}
\allowbreak+\allowbreak\mathcal{F}_\mathrm{xc}$.
In doped semiconductors $\mathcal{F}_\mathrm{xc}$ has negligible influences on both the stationary and dynamic properties of the electrons \cite{Shojiro1979}. Hereafter we set $\mathcal{F}_\mathrm{xc}=0$. The generalized Thomas-Fermi energy density is expressed as (Supplementary Note~2)
\begin{equation}
\label{Eq:FreeEnergyDensityOfTF}
\mathcal{F}_\mathrm{TF}[n]
= \frac{n\eta(n)}{\beta}-\frac{1}{\beta}
\int_{0}^{n}\!\!n^\prime\,\frac{\mathrm{d}\eta(n^\prime)}{\mathrm{d} n^\prime}\,\mathrm{d}n^\prime.
\end{equation}
Here $\eta(n(\mathbf{r}))$ is the effective local Fermi level and bijectively related to the electron density via 
\begin{equation}
    n(\mathbf{r})=\frac{1}{\pi^{2}}\int_0^{\infty}\!\mathrm{d}k\, k^2f_\mathrm{FD}(\eta(\mathbf{r});E_\mathrm{c}(k)),
\end{equation}
with
\begin{equation}
    f_\mathrm{FD}(\eta(\mathbf{r});E_\mathrm{c}(k))=\frac{1}{\exp[\beta E_\mathrm{c}(k)-\eta(\mathbf{r})]+1}
\end{equation}
being the position-dependent Fermi-Dirac distribution function. 
One can numerically obtain the relation $\eta(n)$ for a specified $E_\mathrm{c}(k)$ (see Methods).
The generalized von Weizs\"{a}cker energy density is given by (see Supplementary Note~2)
\begin{equation}
\label{Eq:FreeEnergyDensityOfVW}
\mathcal{F}_\mathrm{W}[n]
= \lambda_\mathrm{W}\,\zeta(n)(\nabla n)^2.
\end{equation}
Here the nonlocal function $\zeta(n)$ reads 
\begin{equation}
	\zeta(n) = 
	\frac{1}{\,8\pi^2}\,
	\Big[\frac{\mathrm{d}\eta(n)}{\mathrm{d}n}\Big]^2\!
	\int_{0}^{\infty}\!\!\!\!\mathrm{d}k\,\,
	\widehat{\mathcal{O}}[f_\mathrm{FD}]\,E_\mathrm{c}(k),
	\label{Eq:ZetaFunc}
\end{equation}
where 
\begin{equation}
\widehat{\mathcal{O}}[f_\mathrm{FD}]
=[
(2+k\partial_k)(\partial_\eta^2f_\mathrm{FD})
\allowbreak+\allowbreak
(\partial_\eta^2f_\mathrm{FD})\,k\partial_k]\,
k\partial_k
\end{equation}
is an $f_\mathrm{FD}$-dependent operator acting on $E_\mathrm{c}(k)$. 
The von Weizs\"{a}cker parameter $\lambda_\mathrm{W}$ is set to the commonly used value of $1/9$ for optical responses (see Supplementary Note~4). As discussed later, $\lambda_\mathrm{W}$ becomes dependent on the doping desity for stationary states. Notably, $\eta(n)$ and $\zeta(n)$ encode the effects of the bound electrons and atomic lattices on conduction electrons via nontrivial dispersion $E_\mathrm{c}(k)$.
Eqs.~\eqref{Eq:FreeEnergyDensityOfTF} and \eqref{Eq:ZetaFunc} form the generalization of the finite-temperature Thomas-Fermi and von Weizs\"{a}cker energies for arbitrary isotropic energy-momentum dispersions. 
After obtaining free energy density $\mathcal{F}[n]$, we can express the functional derivative of the free energy $F[n]$ as
\begin{align}
\label{Eq:FunctionalDerivative}
\frac{\delta F[n]}{\delta n}
=
\frac{\eta(n)}{\beta}
-\lambda_\mathrm{W}\!\Big[
\frac{\mathrm{d}\zeta(n)}{\mathrm{d}n}(\nabla n)^2 
+ 2\zeta(n)\nabla^2n
\Big].
\end{align}

\subsection{Restoring operator $\widehat{\Theta}_\mathrm{R}$ and damping operator $\widehat{\Gamma}_\mathrm{D}$}
The explicit form of the restoring operator $\widehat{\Theta}_\mathrm{R}$ can be obtained by substituting Eq.~\eqref{Eq:FunctionalDerivative} into Eq.~\eqref{Eq:RestoringForce}. It is
\begin{equation}\label{Eq:ThetaRExpression}
    \widehat{\Theta}_\mathrm{R}\mathbf{X}
\!=\! \rho \nabla [g_1(n_0) \nabla\!\cdot\!(\rho \mathbf{X})]\! +\! \rho\nabla\nabla [g_2(n_0) \nabla\nabla\!\cdot\!(\rho \mathbf{X})],
\end{equation}
with $g_1(n_0)=-\frac{n_+}{\beta}\Difft{\eta(n_0)}{n_0}+\lambdaWRes \nion [\nabla^2\zeta(n_0) + \Diff{\zeta(n_0)}{n_0}\nabla^2 n_0]$ and $g_2(n_0)$ $=$ $2\lambdaWRes \nion\zeta(n_0)$.
As indicated by Eq~\eqref{Eq:ThetaRExpression}, the restoring force $\widehat{\Theta}_\mathrm{R}\mathbf{X}$ in the generalized Lorentz model Eq~\eqref{Eq:GeneralizedLorentzByDisplacement} incorporates the nonlocal responses, arising from the inhomogeneities of both the stationary carrier density $n_0(\mathbf{r})$ and the carrier displacement $\mathbf{X}$. 

The damping comprises a phenomenological part with a rate $\gamma_\mathrm{D}$ and a significant nonlocal damping described by a shear viscous tensor $\sigma_\mathrm{L}[\dot{\mathbf{X}}]$ \cite{cirac2017CurrentDependent,zhang_quantum-hydrodynamic_2025}. We have
\begin{equation}
\label{Eq:DampingForce}
\widehat{\Gamma}_\mathrm{D}\dot{\mathbf{X}}
=\rho m^* \gamma_\mathrm{D}\dot{\mathbf{X}}
- \nabla\cdot\sigma_\mathrm{L}[\dot{\mathbf{X}}]/n_+,
\end{equation}
with
\begin{equation}
\label{Eq:ViscosTensorForDamping}
[\sigma_\mathrm{L}[\dot{\mathbf{X}}]]_{ij} = 
\lambda_\mathrm{L}f_\mathrm{CV}(n_0)\!\Big(
\partial_j\dot{X}_i+\partial_i\dot{X}_j
-
\frac{2}{3}\delta_{ij}\nabla\!\cdot\!\dot{\mathbf{X}} 
\Big), 
\end{equation}
where $\dot{\mathbf{X}}=\partial_t\mathbf{X}$ and $f_\mathrm{CV}(n)$ is the Conti-Vignale interpolation function \cite{Vignale_elasticity_1999}.
Here the subscripts $i$ and $j$ label the Cartesian components and $\partial_i$ denotes the partial derivative with respect to the $i$-th spatial coordinate. We find that the preceding scaling factor $\lambda_\mathrm{L}$ is doping dependent and has the following empirical interpolatory formula:
\begin{align}
	\label{eq:CoefficientLD}
\lambda_\mathrm{L}(\overline{\mu})
= 17.1 
+ 55.5\overline{\mu}
+ 293e^{-81.9\overline{\mu}}.
\end{align}
As shown in Supplementary Note~4, the formula is obtained by comparing the computed plasmon linewidths for semiconductor nanospheres with the results from local random phase approximation \cite{weick_surface_2006}. The doping dependence in nonlocal damping is a unique feature for semiconductors.

\subsection{Generalized Lorentz framework and Modal analysis} Equation \eqref{Eq:GeneralizedLorentzByDisplacement} and the Maxwell equations constitute a generalized Lorentz framework (GLF) for studying active semiconductor plasmonics at the deep nanoscale. Its frequency-domain formulation using $\exp(-i\omega t)$ reads
\begin{align}
\label{Eq:OpticalResponseP}
\omega^2\mathbf{P}
+ i\omega\frac{\rho}{\,m^*}\widehat{\Gamma}\mathbf{P}
- \frac{\rho}{\,m^*}\widehat{\Theta}\mathbf{P}
&= -\varepsilon_0\omega_\mathrm{p}^2\rho\mathbf{E},
\\[3pt]
\label{Eq:OpticalResponseE}
\nabla{\times}\nabla{\times}\mathbf{E}
-\varepsilon_0\varepsilon_\infty\mu_0\omega^2\mathbf{E}
&= \mu_0\omega^2\mathbf{P},
\end{align}
with polarization
$\mathbf{P}
\!\allowbreak=\!\allowbreak
-en_0\mathbf{X}$
and 
$\omega_\mathrm{p}
\!\allowbreak=\!\allowbreak
[e^2n_+/(\varepsilon_0m^*)]^{1/2}$.
$\mu_0$ and $\varepsilon_\infty$ 
are the vacuum permeability and high-frequency permittivity, respectively.
The expressions of $\widehat{\Gamma}
\allowbreak\!=\!\allowbreak
\rho^{-1}\widehat{\Gamma}_\mathrm{D}\rho^{-1}$
and
$\widehat{\Theta}
\allowbreak=\allowbreak
\rho^{-1}\widehat{\Theta}_\mathrm{R}\rho^{-1}$ follow straightforwardly from Eqs.~\eqref{Eq:ThetaRExpression} and \eqref{Eq:DampingForce}, and read
\begin{align}
    \widehat{\Theta} =& \nabla [g_1(n_0) \nabla\cdot] + \nabla\nabla [g_2(n_0) \nabla(\nabla\cdot)] ,
    \\
    [\widehat{\Gamma}]_{j\jj} =& \frac{\mResp\gammaD}{\rho}\delta_{j\jj} - \sum_{ii^\prime}\frac{ \delta_{\ii j} \delta_{i\jj}  +\delta_{\ii i} \delta_{j\jj} - \frac{2}{3} \delta_{ij}\delta_{\ii\jj}}{n_+}  \times
    \notag \\
    &\frac{1}{\un}\partial_i\Big[\etaLD f_\text{CV}(n_0) \partial_{\ii} \Big(\frac{1}{\un}\cdot\Big)\Big],
\end{align}
with the $\delta_{ij}$ being the Kronecker delta. Both $\widehat{\Theta}$ and $\widehat{\Gamma}$ are symmetric operators (see Supplementary Note~4).

Based on the GLF, we formulate the source-free Maxwell equations into a linear eigenvalue problem
\begin{equation}\label{eq:QNMMatrix}
    \widehat{H}\psiv \!\equiv \!\begin{pmatrix}
        0                            & \frac{i}{\epso\epsinf}\nabla\times & 0               & -\frac{i}{\epso\epsinf} \\
        -\frac{i}{\mu_0}\nabla\times &                                  0 & 0               & 0 \\
        0                            &                                  0 & 0               & i \\
        i\un\epso\omegaP^2           &                                  0 & -i\frac{\un}{\mResp}\ThetaMat & -i\frac{\un}{\mResp}\GammaMat 
    \end{pmatrix}\!\psiv\!
    =\!\omega\psiv,
\end{equation}
where $\Psi = [\mathbf{E}, \mathbf{H}, \mathbf{P}, \mathbf{J}]$. 
Consequently, QNMs can be defined according to $\widehat{H}\Psi_n=\widetilde{\omega}_n\Psi_n$,with $\widetilde{\omega}_n$ and $\Psi_n$ being the $n$th complex eigenfrequency and quasinormal mode, respectively.
Crucially, the symmetry property of $\widehat{\Theta}$ and $\widehat{\Gamma}$ in the GLF allows us to establish a rigorous bi-orthonormal relation,
\begin{equation}
((\Psi_m,\Psi_n))_{\mathcal{M}}
=\int_{\Omega}\rmd{\rv} \psiv_m^\trans\mathcal{M} \psiv_n 
=\delta_{mn},
\end{equation}
with the integral kernel $\mathcal{M}\!=\!\mathrm{diag}\{\epso\epsinf,\allowbreak -\mu_0,\allowbreak \frac{1}{\epso\mResp\omegaP^2}\ThetaMat, \allowbreak -\frac{1}{\epso\omegaP^2\un}\}$
for the QNMs, eliciting a canonical modal expansion formula for the optical responses \cite{sauvan_normalization_2022} (Supplementary Note~5). 
This enables the understanding of the complex optical responses in terms of a few modes. 

\section{Results}
\subsection{Self-consistent stationary electron distribution}
\label{sec:StationaryElectron}
One of the key features of our GLF is that the stationary electron distribution $n_0(\mathbf{r})$ can be obtained self-consistently.
By inserting the functional derivative obtained in Eq.~\eqref{Eq:FunctionalDerivative}, Eq.~\eqref{Eq:GeneralizedLorentzStationary} yields
\begin{equation}
	\label{Eq:EtaExpression}
	\frac{\eta(n_0)}{\beta}\!=\!\mu\!+\!e\phi_0 
	\!+\!\lambda_\mathrm{W}\Big[\frac{\mathrm{d}\zeta(n_0)}{\mathrm{d}n_0}(\nabla n_0)^2\!+\!2\zeta(n_0)\nabla^2n_0\Big],
\end{equation}
revealing that $\eta/\beta$ differs from the classical local Fermi level $\mu+e\phi_0$ by a nonlocal (gradient) correction arising from the generalized von Weizs\"{a}cker energy functional, which enables our framework to capture nontrivial stationary electron distributions and quantum (nonlocal) effects beyond classical local theory.
The stationary density $n_0$ can be determined by solving Eq.~\eqref{Eq:EtaExpression} together with the Poisson equation 
\begin{equation}\label{eq:Poisson}
\nabla\cdot(\varepsilon_\mathrm{s}\nabla\phi_0)=\frac{e(n_0-n_+)}{\varepsilon_0},
\end{equation}
where $\varepsilon_0$ and $\varepsilon_\mathrm{s}$ are the vacuum and static permittivity, respectively. $\lambda_\mathrm{W,s}$ for stationary state is susceptible to electron interaction characterized with chemical potential $\mu$ which depends on the doping density $n_+$ and $E_\mathrm{c}(k)$ (see Eq.~\eqref{Eq:defChemicalPotential}).
We have obtained an empirical interpolatory formula as (Supplementary Note~3)
\begin{equation}\label{eq:LambdaWEquilibrium}
	\lambda_\mathrm{W,s}(\overline{\mu})
	= 1/9
	+ 0.502\,e^{8.68\overline{\mu}}/(1.61 + e^{8.60\overline{\mu}}),
\end{equation}
with $\overline{\mu}\!=\!\mu/\mathrm{eV}$.
Given $n_0$, we obtain the inhomogeneous effective mass 
\begin{equation}
	m^*(\mathbf{r})=\left.\left[\hslash^2k\left(\frac{\partial E_\mathrm{c}(k)}{\partial k}\right)^{-1}\right]\right|_{E_c=\eta(\mathbf{r})/\beta}
\end{equation} 
evaluated at the effective local Fermi level and corresponding Fermi wavevector (Supplementary Note~2).

\begin{figure}[!t]
\centering
\includegraphics[width=\linewidth]{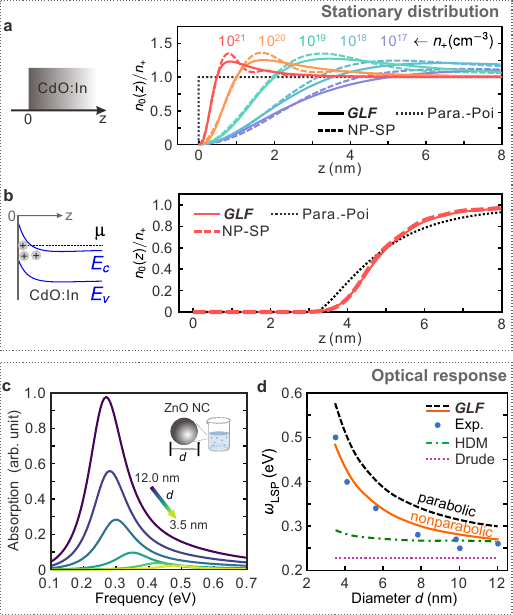}
\caption{\textbf{Validation of the developed theory.} \textbf{a}, GLF, NP-SP, and Para-Poi theories are deployed for computing $n_0$ distributions near the CdO:In/vacuum surface in the neutral case. \textbf{b}, Same as \textbf{a} for the charge-depletion case, with surface charge density $\sigma\allowbreak\!=\!\allowbreak 5\!\times\!10^{14}$ cm$^{-2}$ and 
$n_+\allowbreak\!=\!\allowbreak1{\times}10^{21}$ cm$^{\!-3}$.  
\textbf{c}, Calculated absorption spectra for photodoped ZnO nanocrystals ($n_+=1{\times}10^{20}$~$\text{cm}^{-3}$) of varying sizes.
\textbf{d}, Resonance frequencies (extracted from \textbf{c}) versus the nanocrystal diameter. Experimental data are from Ref.\,\cite{schimpf_charge-tunable_2014}.
The dashed trace denotes calculation without considering the nonparabolicity. 
The green dash-dotted line represents the result obtained by the HDM with an adapted metal-form free-energy functional.}
\label{fig:fig2}
\end{figure}

\subsection{Validation of the theory}
We firstly validate the GLF for evaluating the stationary electron distribution with a 1D CdO:In/vacuum structure where a rigorous quantum description based on the nonparabolic Schr\"{o}dinger-Poisson (NP-SP) theory \cite{abe_evolution_2002} can be used (see Supplementary Note~6). 
The conduction electrons in CdO:In exhibits a nonparabolic dispersion \cite{mendelsberg_determining_2012}: $\hslash^2k^2/(2m^*_0)\!=\!E_\mathrm{c}+C\!E_\mathrm{c}^2$, with $C\!=\!0.5$ eV$^{-1}$ and $m^*_0\!=\!0.16\,m_e$, where $m_e$ is the electron mass. 
CdO:In is reported to sustain heavy doping up to $10^{21}$ cm$^{\!-3}$ \cite{giannuzzi_tunable_2019}. Here we examine $n_0$ near the interface for doping densities spanning five orders of magnitude, i.e., from $10^{17}$ cm$^{\!-3}$ to $10^{21}$ cm$^{\!-3}$. 
Fig.\,\ref{fig:fig2}a presents the calculated $n_0$ versus position, with solid traces denoting results from the GLF and dashed traces from the NP-SP method. 
Across all doping densities, the two approaches exhibit good agreement, showing a transition behavior of $n_0$ with a doping-dependent depletion region near the interface. 
The popular classical parabolic-Poisson (Para-Poi) theory (see Supplementary Note 6) is also deployed for comparison. It completely misses the transition behavior. 

Charging could affect
band bending and the depletion or accumulation of electrons near the interface, leading to an effective surface charge density $\sigma$ as shown on the left of Fig. \ref{fig:fig2}b. 
Assuming $\sigma\allowbreak\!=\!\allowbreak 5\!\times\!10^{14}$ cm$^{-2}$ and a doping density of $n_+\allowbreak\!=\!\allowbreak10^{21}$ cm$^{-3}$, as presented on the right of Fig.\,\ref{fig:fig2}b, results from our GLF and the NP-SP method completely overlap in the entire region, yielding a depletion layer of $4$ nm. 
In contrast, results from the parabolic-Poisson approach (dotted line) show significant deviations. 
Besides the benchmarks for depletion layers, further validation concerning charge accumulation layers in the narrow-gap semiconductors InAs and InSb is provided in Supplementary Note~7.
Correct evaluation of $n_0$ is crucial as it affects the plasmonic responses and their susceptibility to electrical control. 

Next, we demonstrate the validity of the GLF for the optical response by examining the absorption spectra of ZnO nanocrystals sized from 3.5 nm to 12 nm as shown in Fig.\,\ref{fig:fig2}c. 
The experimental data are available from Ref.\,\cite{schimpf_charge-tunable_2014}.
As evident in Fig.\,\ref{fig:fig2}d, calculations from our GLF incorporating conduction-band nonparabolicity~\cite{kim_density_state_2008} align closely with the experimental observations. Ignoring conduction-band nonparabolicity leads to systematically overestimated resonance frequencies. 
It might be also tempting to see the results obtained by applying the metal-form free energy functional with the use of semiconductor's doping density and effective mass. As shown by the green dash-dotted line in Fig.\,\ref{fig:fig2}d, the results fail to reproduce the expected nonlocal resonance shift as the particle size shrinks. Moreover, the widely-used Drude model completely misses out the size effect on plasmon resonances.
In Supplementary Note~8, we show the agreements between GLF results and experimental data for CdO:In nanocrystals \cite{shubert-zuleta_how_2024}.


\subsection{Electro-plasmon modulation in a semiconductor nanoantenna} 
\begin{figure}[!t]
\centering
\includegraphics[width=\linewidth]{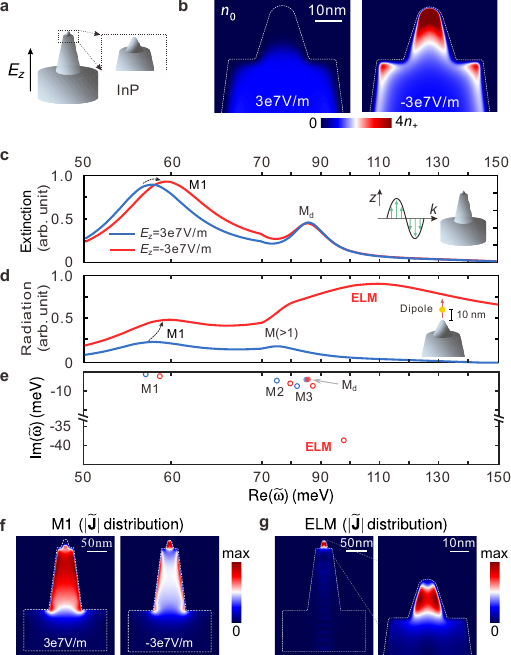}
\caption{\textbf{Electro-plasmon modulation with an InP nanoantenna.} \textbf{a}, Schematics of the nanoantenna structure. \textbf{b}, Distributions of $n_0$ at $E_z\allowbreak\!=\!\allowbreak\pm3{\times}10^7$ V/m. \textbf{c}, Extinction spectra under plane-wave excitation. \textbf{d}, Radiation rate spectra of an electric dipole. \textbf{e}, QNMs on the complex-frequency plane. Modal current distributions for modes \textbf{f}, M1 and \textbf{g}, ELM.}
\label{fig:fig3}
\end{figure}

As an exemplary application, we study electro-plasmon modulation with an InP nanoantenna. 
As sketched in Fig.~\ref{fig:fig3}a, it consists of a disk substrate (diameter $200$ nm, height $100$ nm), a 150-nm-high cone frustum, and a 20-nm-tall protrusion on the top (see Methods for modeling details). 
Such structures are experimentally feasible with state-of-the-art techniques \cite{bakkers_epitaxial_2004,mcintyre_semiconductor_2020}. We consider a doping density of $n_+\!=\!5\times10^{18}$ cm$^{\!-3}$ and perform GLF calculations taking the complicated $E_\mathrm{c}(k)$ of InP into account (see Methods).
Fig.~\ref{fig:fig3}b presents the stationary electron distributions under electrostatic fields. 
At $E_\mathrm{z}\allowbreak\!=\!\allowbreak3{\times}10^7$ V/m, electrons are completely depleted in the protrusion, whereas at $E_\mathrm{z}\allowbreak\!=\!\allowbreak-3{\times}10^7$ V/m they accumulate with a density exceeding $4n_+$. 
Since the plasmon frequency scales with the carrier density, we anticipate strong electrical responsivity of the plasmon resonances.

We compute the optical response of the nanoantenna under electrical control for two types of excitations: plane-wave (far-field) and dipole-source (near-field) excitations. Fig.~\ref{fig:fig3}c shows the extinction spectra under plane-wave excitation, whereas Fig.~\ref{fig:fig3}d presents the radiation-rate spectra of an electric dipole placed 10 nm above the protrusion. In both cases, the spectra undergo pronounced changes with the electrostatic field, although the detailed features differ substantially. Under plane-wave excitation, the resonance around $60$~meV exhibits a sizable shift, while the peak around $85$~meV remains largely unchanged. For dipole excitation, the radiation spectra reveal an additional broad resonance, indicating intriguing physics.

To understand the physics, we perform modal analysis. Fig. \ref{fig:fig3}e displays the eigenfrequencies of the involved QNMs on the complex-frequency plane, where the real eigenfrequency represents the resonance position while the imaginary part indicates the resonance linewidth. 
The blue and red circles denote the modes under  $E_\mathrm{z}\allowbreak\!=\!\allowbreak-3{\times}10^7$ V/m and $E_\mathrm{z}\allowbreak\!=\!\allowbreak3{\times}10^7$ V/m, respectively. 
One sees that the resonance around $60$~meV is caused by QNM M1 while the emergence of the broad resonance in Fig.~\ref{fig:fig3}d is due to the formation of a new eigenmode denoted as ELM. 
The secondary peak near $85$~meV in Fig.~\ref{fig:fig3}c is due to the disk mode $\text{M}_\text{d}$ (Supplementary Note~10), which is thus insensitive to electrical tuning. 
The pronounced shift of mode M1 can be better understood by examining its modal current $\mathbf{J}$ distributions depicted in Fig.~\ref{fig:fig3}f, where the profile changes dramatically with the applied field. 
The formation of ELM is due to electron accumulation in the protrusion for $E_\mathrm{z}\allowbreak\!=\!\allowbreak-3{\times}10^7\,\mathrm{V/m}$. 
The modal current $\mathbf{J}$ displayed in Fig.~\ref{fig:fig3}g induces an extremely localized modal field with a mode volume of $5.6\times10^3\,\mathrm{nm}^3$ \cite{Li2021,Wu2021,benz_single-molecule_2016} (Supplementary Note~10). 
Such confinement triggers severe nonlocal damping and thus the resonance quality factor is only about 1.3. When switched to $E_\mathrm{z}=3{\times}10^7$ V/m, electrons in the protrusion are depleted, and the ELM stunningly vanishes. By involving carrier dynamics only in a tiny structure, such electrical modulation promises an ultrafast operation.

\section{Conclusion}
We have developed a generalized Lorentz framework for active semiconductor surface nanoplasmonics. 
The framework captures key nonclassical effects, including carrier nonlocality, band bending near the interface, doping-density-dependent von Weizs\"{a}cker functional and nonlocal damping. 
Specifically, for treating carrier nonlocality, we generalize the finite-temperature Thomas-Fermi and von Weizs\"{a}cker energy functionals for arbitrary isotropic energy-momentum dispersion relations (Eqs.\,\eqref{Eq:FreeEnergyDensityOfTF}-\eqref{Eq:ZetaFunc}). 
These functionals are the cornerstone of various mean-field theories for many-body electron systems \cite{Feynman1949,Kirzhnits1956,Perrot1979}.
Our framework allows efficient computation of the carrier dynamics and optical responses for multiscale 3D semiconductor nanostructures. Moreover, this framework enables a rigorous quasinormal mode analysis, providing a transparent modal interpretation of complex optical responses. We validate the theory through comparisons with Schr\"odinger-Poisson calculations for 1D planar structures and with experimental observations for semiconductor nanocrystals of varying sizes. Further, we apply the framework to describe electro-plasmon modulation in a semiconductor nanoantenna and demonstrate that the electrically induced local depletion and accumulation of electrons can create or annihilate the extremely localized eigenmode.

Caution should be exercised with the applicability when extreme damping due to excessive doping extinguishes the plasmon or the structure size is smaller than the ballistic length of electrons (cf.~Supplementary Note~9 for a detailed evaluation). The present formulation is extendable---for example, to incorporate anisotropic band dispersion \cite{Shokhovets2008.PRB}, $p$-doped semiconductors \cite{Maack2018.PRB}, heterostructures, and exchange-correlation effects in strongly correlated regimes \cite{Gammon1992.PRL}. Our work establishes a comprehensive quantitative framework for active semiconductor quantum surface plasmonics and will foster the convergence of electronics and photonics at the deep nanoscale.

\section{Methods}

Detailed derivations of the generalized Lorentz model, the free-energy functional, the restoring and damping operators, and the quasinormal-mode formalism, together with additional validation results, are provided in the Supplementary Information.

\subsection{Boundary conditions for stationary equations}
The nonlinear equation~\eqref{Eq:GeneralizedLorentzStationary} and the Poisson equation~\eqref{eq:Poisson} form a coupled system for determining the self-consistent stationary electron density $n_0$.
On the boundary of the computational domain $\Omega$ for the stationary electron density $n_0$, a Dirichlet boundary condition $n_0\vert_{\partial \Omega}=0$ is imposed.
Meanwhile, the electrostatic potential $\phi_0$ must satisfy $\oint_{\partial\Omega}\nabla\phi_0\cdot\rmd{\Sv}=0$, if the net charge of the semiconductor is zero.

\textbf{In a uniform electrostatic field.}
We assume the nanostructure is placed in a uniform background electrostatic field $\Evb$, e.g., inside a charged parallel-plate capacitor.
The governing equations~\eqref{Eq:GeneralizedLorentzStationary} and~\eqref{eq:Poisson} for $n_0$ and $\phi_0$ remain unchanged.
The boundary condition for $\phi_0$ becomes $\lim\limits_{\rv\rightarrow\infty}\nabla\phi_0=-\Evb$.

\textbf{Electrically charged semiconductor.}
When a semiconductor nanostructure is electrically charged with net charge $-e\Delta N$, the form of Eqs.~\eqref{Eq:GeneralizedLorentzStationary} and~\eqref{eq:Poisson} remains unchanged, 
but the boundary condition for the electrostatic potential $\phi_0$ must be modified.
The Neumann boundary condition $\Diff{\phi_0}{\hat{\rv}}\vert_{\partial \Omega}$ is imposed to account for the net charge.
We require that the electrostatic field $\mathbf{E}_0=-\nabla\phi_0$ satisfy Gauss's law $\epso\int_{\partial\Omega}\epstat\mathbf{E}_0\cdot d\mathbf{S}=-e\Delta N$.
In practice, we impose a spherical boundary of radius $r_0$ that is sufficiently distant from the nanostructure of characteristic size $a$, such that the field at the boundary can be well approximated as that emanating from a point charge located at the center.
The boundary condition then simplifies to $\epstat\Diff{\phi_0}{r}\vert_{r=r_0}=-\frac{e\Delta N}{4\pi r_0^2\epso}$ ($r_0\gg a$).

\subsection{Nonparabolic energy-momentum dispersion relations}
For all the semiconductors considered in this work, the dispersion relations follow from Kane's $\kv\cdot\pv$ theory \cite{kane_band_1957}.
In general, the energy dispersion is implicitly expressed as
\begin{equation}\label{eq:KaneDispersion1}
    \Ep(\Ep+\Eg)(\Ep+\Eg+\Deltasp) - k^2P^2(\Ep+\Eg+2\Delta_\mathrm{so}/3) = 0,
\end{equation}
where the conduction band minimum (CBM) in the bulk region is set as the zero-energy reference.
Kane's matrix element $P^2$ is
\begin{equation}
    P^2 = \frac{3\hslash^2(1/\mestar - 1/\me)}{2[2/\Eg+1/(\Eg+\Deltasp)]}.
\end{equation}
Here $\mestar$, $\Deltasp$ and $\Eg$ are the effective mass at the CBM, spin-orbit splitting energy and band gap, respectively.
Equation~\eqref{eq:KaneDispersion1} is a cubic equation in $\Ep$ and can be solved by Cardano's formula.
After solving for $\Ep$, the conduction band is expressed as $\EC(k)=\Ep + \frac{\hslash^2k^2}{2\me}$.

For semiconductors like InN we have $\Deltasp\ll\Eg$, so the conduction band simplifies to
\begin{equation}\label{eq:KaneDispersion2}
    \EC(k) = \frac{1}{2} (-\Eg + \sqrt{\Eg^2+4k^2P^2}) + \frac{\hslash^2k^2}{2\me}
\end{equation}
with $P^2 = \frac{\hslash^2}{2\me}\big(\frac{\me}{\mestar}-1\big)\Eg$.
When $\Eg$ is large enough, e.g.~for ITO and CdO:In, the conduction band is further simplified to the implicit form
\begin{equation}\label{eq:KaneDispersion3}
    C\EC^2 + \EC = \frac{\hslash^2k^2}{2\mestar}.
\end{equation}
Table~\ref{tab:NPbandParameters} summarizes the dispersion model parameters for the relevant semiconducting materials in this work.

\begin{table}[!t]
\footnotesize
\setlength{\tabcolsep}{2pt}
    \centering
    \caption{Dispersion model parameters of the conduction band for $n$-doped ITO, CdO, ZnO, InAs, InN and InP. $\Eg$: band gap; $\mestar$: effective mass at the CBM; $\Deltasp$: spin-orbit splitting; $C$: nonparabolic parameter.}
    \label{tab:NPbandParameters}
    \begin{tabular}{lcccccc}
    \toprule
    & ITO\cite{liu_quantification_2014} & CdO\cite{mendelsberg_determining_2012} & ZnO\cite{kim_density_state_2008} & InAs\cite{abe_evolution_2002} & InN\cite{king_nonparabolic_2008} & InP\cite{vurgaftman_band_2001} \\
    \midrule
    $\Eg$~(eV) & -- & -- & -- & 0.3543 & 0.69 & 1.4236 \\
    $\mestar$~($\me$) & 0.263 & 0.16 & 0.26 & 0.02195 & 0.048 & 0.0795 \\
    $\Deltasp$~(eV) & -- & -- & -- & 0.381 & 0 & 0.108 \\
    $C$~(eV$^{-1}$) & 0.4191 & 0.5 & 0.35 & -- & -- & -- \\
    \bottomrule
    \end{tabular}
\end{table}

\subsection{Material parameters of several semiconductors}
\label{Sec:OpticalParameters}
The material parameters for the stationary and optical-response calculations are listed in Table \ref{tab:OpticalParameters} for the materials involved in this work.
{Note that we neglect the anisotropy of ZnO. Since the calculations consider ZnO nanocrystals dispersed in solution with random orientations, the anisotropic contributions largely average out in the ensemble-averaged response.}

\begin{table}[!t]
\footnotesize
\setlength{\tabcolsep}{2pt}
	\centering
    \caption{Material parameters for optical response calculation. $\epstat$: static relative dielectric constant; $\epsinf$: high frequency dielectric constant; $\gammaD$: phenomenological damping rate.}
	\label{tab:OpticalParameters}
    \begin{tabular}{ccccccc}
	\toprule
    {}&
    ITO\cite{liu_electrical_2017}&
    CdO\cite{liu_effects_2016}&
    ZnO \cite{madelung_semiconductors_2004}&
    InAs \cite{madelung_semiconductors_2004}&
    InN \cite{madelung_semiconductors_2004} & 
    InP \cite{madelung_semiconductors_2004}\\
    \midrule
    $\epstat\,(1)$ & 9.3 & 21.9 & 7.8 & 15.15 & 9.3 & 12.56 \\
    $\epsinf\,(1)$ & 3.9 & 5.0 & 3.72 & 12.37 & 6.7\cite{askari_low-loss_2018} & 9.55\\
    $\gammaD\,(\text{eV})$ & 0.2 & 0.059 & 0.13 & 0.001 & 0.012 & 0.011 \\
	\bottomrule
    \end{tabular}
\end{table}

\subsection{Treatment of the integrals in free energy functional}
\label{Sec:ImproperIntegrals}
During the numerical implementation of GLF, the free-energy functional and its functional derivatives involve integrals with infinite limits, $\int_0^\infty\mathrm{d}k\,\{\cdot\}$.
Specifically, the variables containing such integrals include $n_0(\eta)$, $\eta(n_0)$, $\Diff{\eta(n_0)}{n_0}$, $\zeta(n_0)$, $\Diff{\zeta(n_0)}{n_0}$ and $\DDiff{\zeta(n_0)}{n_0}$.
Neither analytical closed-form expressions nor direct numerical implementations are available for these quantities, with the latter hampered by numerical instabilities stemming from nonparabolic Fermi-Dirac integrals.
Owing to the bijective relation between $n_0$ and $\eta$, all these functions can be expressed in terms of $\eta$. Their numerical evaluation is well-behaved when implemented in a piecewise manner.
When $\eta$ is between $-20$ and $150$, these functions can be evaluated accurately using the `scipy.integrate.quad' function from the SciPy library (version 1.15), which adaptively applies the Gauss-Kronrod quadrature rules to subintervals.
The resulting integral values at the sampled $\eta$ points are then used to construct interpolation functions for the numerical model.
This range covers the common electron concentrations from $1\times 10^{16}$ to $5\times 10^{21}\,\text{cm}^{-3}$.
In the vicinity of the semiconductor boundary, the electron density $n_0$ rapidly decays to zero, and $\eta$ tends to $-\infty$.
In this low-density regime ($\eta<-20$), accurate evaluation via numerical integration becomes difficult, and we employ asymptotic analytic formulae instead.
Taking $n_0(\eta)$ as a representative case, we present the derivation of the asymptotic analytic formula.
The asymptotic analytic formulae of the other functions will be provided without proof.

\begin{table}[!t]
\footnotesize
\setlength{\tabcolsep}{2pt}
    \caption{Numerical implementation of functions containing integrals with infinite limits. 
    For $\eta\in(-20,150]$, the tabulated integrals are directly evaluated with numerical integration and serve for interpolation function construction. 
    For $\eta\in[-\infty,-20]$, the functions are accurately approximated by asymptotic analytic expressions.
    The `const.' entries are determined numerically.}
	\label{tab:functionlimit}%
    \begin{tabular}{ccc}
		\toprule
		&
		$\eta\in(-20,150]$&
		$\eta\in[-\infty,-20]$\\
		\midrule
		$n_0(\eta)$ & $\frac{2}{(2\pi)^3}\int\rmd{\mathbf{k}}\frac{1}{\exp(\beta \EC(k)-\eta(\rv))+1}$ & const.$\times e^\eta$ \\
		$\eta(n_0)$ & $n_0^{-1}(\eta)$ & const.$+\ln(n_0)$ \\
		$\Diff{\eta(n_0)}{n_0}$ & $(\Diff{n_0}{\eta})^{-1}$ & const.$/n_0$ \\
		$\zeta(n_0)$ & $\frac{1}{8\pi^2}\Big[\Difft{\eta_{0}}{n_{(0)}}\Big]^2\int_{0}^\infty\mathrm{d}k \Big\{\widehat{\mathcal{O}}[\fFD] \EC \Big\} $ & const.$/n_0$ \\
		$\Diff{\zeta(n_0)}{n_0}$ & $\Diff{\zeta}{\eta}(\Diff{n_0}{\eta})^{-1}$ & const.$/n_0^2$ \\
		$\DDiff{\zeta(n_0)}{n_0}$ & $\Diff{\zetaP}{\eta}(\Diff{n_0}{\eta})^{-1}$ & const.$/n_0^3$ \\
		\bottomrule
    \end{tabular}
\end{table}

The relation between $\eta$ and $n_0$ is implicitly given by the inverse of the improper integral
\begin{align}\label{eq:ImproperIntegral-eta1}
    n_0(\eta) =& \frac{2}{(2\pi)^3}\int\rmd{\mathbf{k}}\frac{1}{\exp(\beta \EC(k)-\eta(\rv))+1} \notag \\
    =& \int_0^\infty \frac{\rmd{k}}{\pi^2}\frac{1}{\exp(\beta \EC(k)-\eta(\rv))+1}.
\end{align}
In the limiting case $\eta(\rv)\rightarrow-\infty$, $e^{-\eta}\rightarrow\infty$, the integrand $\tfrac{1}{\exp(\beta \EC(k)-\eta(\rv))+1}$ tends to $\tfrac{1}{\exp(\beta \EC(k)-\eta(\rv))}$ and the expression for the electron density becomes
\begin{equation}
    \lim\limits_{\eta\rightarrow-\infty}n_0(\eta) 
    = e^{\eta}\int_0^\infty \frac{\rmd{k}}{\pi^2}e^{-\beta \EC(k)}.
\end{equation}
The integral therein is easy to evaluate through numerical integration when the explicit form of $\EC(k)$ is specified.
Therefore, we generally have $n(\eta)=\text{const.}\times e^\eta$ in the range $(-\infty,-20)$.

Table \ref{tab:functionlimit} lists the piecewise implementation for all the functions containing integrals with infinite limits, including the integrals to be evaluated numerically for intermediate $\eta$ range and the asymptotic expressions for $\eta$ approaching $-\infty$.
All the constants in the table are numerically derived in a similar way as exemplified for $n(\eta)$.
Through numerical verification, we confirm that within the full range $\eta\in[-\infty, 150]$, the numerical implementation method guarantees high accuracy (at least 10 significant digits).

\subsection{Detailed geometry, meshing strategy and performance for structure in Fig.~3}
\label{Sec:MeshandEff}
\begin{figure}[!htpb]
    \centering
    \includegraphics[width=\linewidth]{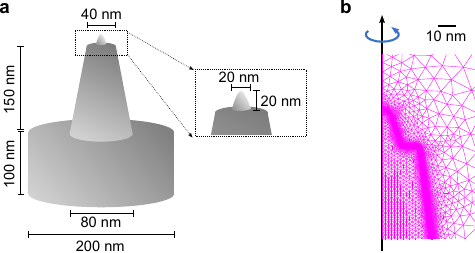}
    \caption{\textbf{Meshing strategy for the InP nanoantenna.} \textbf{a}, Schematic of the nanoantenna geometry as shown in Fig.~3 of the main text. \textbf{b}, The zoomed-in view of the meshing scheme near the protrusion. The maximum mesh size is set in different regions: $1$~nm inside the semiconductor, $0.02$~nm on the boundary, and $20$~nm in the exterior.}
    \label{fig:mesh}
\end{figure}
We take the structure of Fig. 3 as an example to illustrate the meshing strategy and computational performance of solving the GLF with finite-element method.
As shown in Fig.~\ref{fig:mesh}b, the computational domain is divided into three regions.
The maximum mesh size in the internal region of the semiconductor and the external region is set to $1$ nm and $20$ nm, respectively.
On the boundary, the maximum mesh size is set to $0.02$ nm.
All the calculations are performed on a workstation with an `Intel Xeon W9-3475X (36 Cores)' and 256 GB memory.
When using an `Intel oneAPI Math Kernel Library PARDISO' solver, the response calculation takes about $1$ minute and $40$ seconds per frequency point.


\backhead{Data availability}
The data that support the findings of this study are available in the main text and Supplementary Information. Further information is available from the corresponding author upon reasonable request.

\backhead{Code availability}
The codes that support the findings of this study are available from the corresponding author upon reasonable request.

\backhead{Acknowledgements}
We acknowledge financial support from the National Natural Science Foundation of China (12404429, 62235006, 12525414, 12274160, 12504443, 12574416), the China Postdoctoral Science Foundation (2024M760993), the China Postdoctoral Science Foundation-Hubei Joint Support Program (2025T004HB) and the Fundamental Research Funds for the Central Universities (2023BR003) through Huazhong University of Science and Technology. The computations are completed in the HPC Platform of Huazhong University of Science and Technology.

\backhead{Competing interests}
The authors declare no competing interests.


\bibliographystyle{apsrev4-2}
\bibliography{reference}

\end{document}